\documentclass[preprintnumbers,10pt,nofootinbib]{revtex4}

\usepackage{amsmath,latexsym,amssymb,amsfonts}
\usepackage{graphicx}
\usepackage{bm}
\usepackage{epstopdf}
\usepackage{epsfig}
\usepackage{color}

\begin{document}

\title{\textbf{Threshold of primordial black hole formation in Eddington-inspired-Born-Infeld gravity}}

\author{
{\textsc{Che-Yu Chen$^{a,b,c}$}\footnote{{\tt b97202056{}@{}gmail.com}}}
}

\affiliation{
$^{a}$\small{Institute of Physics, Academia Sinica, Taipei, Taiwan 11529}\\
$^{b}$\small{Department of Physics and Center for Theoretical Sciences, National Taiwan University, Taipei, Taiwan 10617}\\
$^{c}$\small{LeCosPA, National Taiwan University, Taipei, Taiwan 10617}
}

\begin{abstract}
It is believed that primordial black holes (PBHs), if they exist, can serve as a powerful tool to probe the early stage of the cosmic history. Essentially, in the radiation dominated universe, PBHs could form by the gravitational collapse of overdense primordial perturbations produced during inflation. In this picture, one important ingredient is the threshold of density contrast, which defines the onset of PBH formation. In the literature, most of the estimations of threshold, no matter numerically or analytically, are implemented in the framework of general relativity (GR). In this paper, by performing analytic estimations, we point out that the threshold for PBH formation depends on the gravitational theory under consideration. In GR, given a fixed equation of state, the analytic estimations adopted in this paper give a constant value of the formation threshold. If the theory is characterized by additional mass scales other than the Planck mass, the estimated threshold of density contrast may depend on the energy scale of the universe at the time of PBH formation. In this paper, we consider the Eddington-inspired-Born-Infeld gravity as an example. We find that the threshold would be enhanced if the Born-Infeld coupling constant is positive, and would be suppressed for a negative coupling constant. Also, we show explicitly that the threshold depends on the energy scale of the universe at the PBH formation time. This conclusion is expected to be valid for any gravitational theory characterized by additional mass scales, suggesting the possibility of testing gravitational theories with PBHs. 
\end{abstract}

\maketitle



\section{Introduction}

The scrutiny of primordial black holes (PBHs) and their cosmological implications have been an intensive field of research over the last decades. Generically, PBHs would form from the gravitational collapse of overdense primordial density perturbations generated quantum mechanically during inflation. Essentially, quantum fluctuations generated during inflation exit the horizon and become classical before the end of inflation. Then, the comoving horizon grows and these density perturbations reenter the horizon. After horizon reentry, if the overdensity of the overdense region is too large, exceeding a certain threshold, the overdense region would collapse and eventually form PBHs. It can be expected that the whole process of PBH formation and its relevant observables would bring several important information about the history of the early universe \cite{Hawking:1971ei,Carr:2005zd,Polnarev:1986bi}, including the imprints of the inflationary model under consideration. Therefore, PBHs can be an important tool to probe the early universe, such as the extremely small scale perturbations which are not accessible by cosmic microwave background observations and the high energy physics up to grand unified scales \cite{Khlopov:1980mg,Kalashnikov:1983qv}. In addition, PBHs with a certain range of mass could be a possible dark matter candidate \cite{Belotsky:2014kca,Carr:2016drx,Inomata:2017okj,Ballesteros:2017fsr}. Also, the merger events of binary black hole systems observed by LIGO/Virgo collaborations indicate a population of black holes which are slightly too massive such that their origin cannot be well-explained by the current model of stellar evolution. In this regard and considering the possibility of late-time successive merging of PBHs \cite{Belotsky:2018wph}, these massive black holes may originate from PBHs \cite{Sasaki:2018dmp}. The possibility that PBHs may be created from cosmological phase transitions was proposed in Refs.~\cite{Konoplich:1999qq,Rubin:2000dq,Khlopov:2000js,Rubin:2001yw,Khlopov:2004sc}. See Ref.~\cite{Khlopov:2008qy} for a nice review of PBH physics.
 
Based on the above picture of PBH formation, one can determine the observables such as the PBH mass function and abundance, which in principle can be constrained by observations \cite{Carr:2009jm}. In determining these observables, the threshold of the density contrast over which the overdense region would collapse to form PBHs plays a very important role. In the literature, several numerical calculations of the PBH formation threshold have been carried out \cite{Niemeyer:1999ak,Shibata:1999zs,Musco:2004ak,Polnarev:2006aa,Musco:2008hv,Musco:2012au,Young:2019yug,Escriva:2019nsa,Musco:2018rwt}. In Ref.~\cite{Carr:1975qj}, a pioneering analytic estimation of the threshold has been performed by using a very simple argument that the threshold is defined by comparing the proper size of the overdense region and its Jeans length. In \cite{Harada:2013epa}, a more refined analytic estimation has been proposed and it has been shown that the refined analytic estimation agrees with numerical simulations qualitatively as compared with the estimation given in Ref.~\cite{Carr:1975qj}. These analytic estimates are feasible because they are based on the assumption that the overdensity is uniform, i.e., the top-hat profile. Generically, the threshold would depend on the shape of the density profile under consideration \cite{Nakama:2013ica,Germani:2018jgr,Musco:2018rwt,Young:2019osy,Escriva:2019phb,Germani:2019zez}, while in these cases only numerical calculations are feasible. Recently, the threshold of density contrast has been estimated when PBHs have spins \cite{Harada:2017fjm,He:2019cdb}. It has been shown that due to the repulsive property of the centrifugal force, the threshold would increase in the presence of PBH spins. This result suppresses the existence of high spin PBHs, which can be confirmed or falsified if the spin distribution of PBHs is observationally available in the future (see Ref.~\cite{Fernandez:2019kyb} in which the authors examined the possibility that LIGO/Virgo events could have PBH origin using the effective spin parameters). The instability and the threshold corresponding to collapsing scalar fields have been studied in Refs.~\cite{Khlopov:1985jw,Hidalgo:2017dfp}.

It should be emphasized that most works about the PBH formation mechanism and the calculations of the formation threshold are mainly within the framework of Einstein's general relativity (GR). However, there are a lot of modified theories of gravity which change the description of the early universe from the standard big bang cosmology. Some of them are motivated by the ambition to ameliorate the initial big bang singularity. One may also treat these modified theories of gravity as effective theories of a fundamental yet unknown quantum theory of gravity. See Refs.~\cite{Capozziello:2011et,Nojiri:2017ncd} for reviews on modified theories of gravity. 
 
Since physics of PBHs highly depends on the early history of our universe, it is expected that the mechanism of PBH formation also depends on the gravitational theories under consideration. This is similar to the fact that different inflationary models naturally give rise to distinct PBH observables. For example, the PBH formation and the threshold has been investigated under the framework of \textit{ad hoc} bouncing cosmologies \cite{Chen:2016kjx}. In Ref.~\cite{Bhattacharya:2019bvk}, the mass function has been analyzed in non-standard cosmologies in which an intermediate stage with a stiff equation of state appears between the end of inflation and the radiation dominated era. The pair creation rate of PBHs in the $f(R)$ gravity was investigated in Ref.~\cite{Dialektopoulos:2017pgo}. Very recently, the primordial curvature perturbation and its influence on PBH formation has been studied in a subclass of scalar-tensor theories \cite{Vallejo-Pena:2019hgv}.
 
In this paper, we will relax the GR assumption and adopt the analytic approaches developed in Refs.~\cite{Carr:1975qj,Harada:2013epa} to estimate the threshold of density contrast for PBH formation. We will particularly choose the Eddington-inspired-Born-Infeld (EiBI) gravity \cite{Banados:2010ix,BeltranJimenez:2017doy} as an example. This theory reduces to GR in vacuum but deviates from it in the presence of matter fields. Most importantly, the EiBI gravity predicts a completely different history of the early universe as compared with that in GR. The initial big bang singularity can be avoided in different manners according to the sign of the Born-Infeld coupling constant \cite{Scargill:2012kg}. In GR, the threshold obtained from the analytic estimations is constant, given a fixed equation of state in the background universe. In EiBI gravity, on the other hand, we will assume a radiation dominated universe and show that the threshold not only deviates from its GR counterpart, but also depends on the energy scale of the universe at the time of PBH formation. This energy scale dependence results from the existence of an additional mass scale characterizing the theory. We expect that such a dependence would happen for any theory with additional mass scales other than the Planck mass.
 
This paper is outlined as follows. In Sec.~\ref{sec.EOM}, we will start with a brief review of the EiBI gravity. Then we will adopt two analytic approaches to estimate the threshold of the density contrast within the EiBI scenario. In Sec.~\ref{sec.mfa}, we will discuss how the PBH mass function and abundance are affected through the changes of the threshold in different gravitational theories. We finally draw our conclusions in Sec.~\ref{sec.conclu}.

\section{Threshold of PBH formation in EiBI gravity}\label{sec.EOM}

In order to estimate the threshold $\delta_{\textrm{th}}$ of the density contrast for PBH formation, we first assume that the overdense region has uniform overdensity and can be approximately described by a part of the closed Friedmann universe \cite{Carr:1975qj,Harada:2013epa}. The metric describing the overdense region reads
\begin{equation}
ds^2=-dt^2+a^2(t)\left(d\chi^2+\sin^2\chi d\Omega_2^2\right)\,,\label{closeFLRW}
\end{equation}
where the scale factor $a(t)$ is a function of the cosmic time $t$. In this setup, the overdense region corresponds to the region where $0\le\chi\le\chi_a$. The areal radius of the overdense region is given by
\begin{equation}
R_a\equiv a\sin\chi_a\,.
\end{equation}
The proper size of the overdense region is assumed to be on super-Hubble scales initially. With the expansion of the universe, the overdense region would reenter the Hubble horizon at the time when the areal radius of the overdense region is equal to the Hubble horizon $1/H$ (in the uniform Hubble slicing):
\begin{equation}
\frac{1}{H_{\textrm{hc}}}=a_{\textrm{hc}}\sin\chi_a\,,\label{reentryxa}
\end{equation}
where the subscript ``hc" stands for quantities at the time of horizon reentry. In addition, the overdense region itself would undergo a transition from the expansion phase to the contraction phase. The scale factor reaches $a_{\textrm{m}}$ at the maximum expansion time. From now on, we will use the subscript ``m" to represent quantities at the maximum expansion time.

In Refs.~\cite{Carr:1975qj,Harada:2013epa}, the threshold $\delta_{\textrm{th}}$ for PBH formation in the GR framework was estimated based on two seemingly similar, but conceptually different arguments, respectively. In Ref.~\cite{Carr:1975qj}, the threshold was obtained from a simple argument that the proper size of the overdense region at the maximum expansion time should be larger than the Jeans length in order to initiate the gravitational collapse. Assuming that the universe is dominated by a perfect fluid with a constant equation of state $w$, the estimated threshold is $\delta_{\textrm{th}}=w$ \cite{Carr:1975qj}. On the other hand, a more refined approach was proposed in \cite{Harada:2013epa} and the threshold was estimated by requiring that the sound crossing distance by the maximum expansion time should be smaller than the proper size of the overdense region for the onset of the collapse. The threshold was estimated as $\delta_{\textrm{th}}=\sin^2\left[\pi\sqrt{w}/(1+3w)\right]$ and it was shown in \cite{Harada:2013epa} that this estimation is in much better agreement with the numerical simulations as compared with the estimation given in Ref.~\cite{Carr:1975qj}. In fact, as clearly explained in Ref.~\cite{Musco:2018rwt}, the estimation in \cite{Harada:2013epa} represents a lower bound of the threshold since even though the estimation takes into account the relativistic gravitational effect of the pressure, the effects contributed by the pressure gradients are not considered. In Ref.~\cite{Carr:1975qj}, the estimation is based on a Newtonian approximation and the result is less accurate. In the following subsections, we will use the two analytic estimations mentioned above in different gravitational theories. As a specific example, we will consider the EiBI gravity and show how the threshold value is affected by different descriptions of the early universe. It should be noted that in the following estimations, we will consider the threshold in the uniform Hubble slicing, which is also used in Refs.~\cite{Carr:1975qj,Harada:2013epa}. As opposed to the comoving slicing which is typically used in numerical simulations, the thresholds in these two slicings are different by a numerical factor depending on the equation of state.

\subsection{EiBI gravity: A brief review}
Before starting the estimation, we will briefly review the EiBI gravity and introduce the modified Friedmann equation describing a closed universe, which is interpreted as the overdense region embedded in a background flat universe. The action of the EiBI gravity \cite{Banados:2010ix,BeltranJimenez:2017doy} is given by{\footnote{In the original EiBI proposal, there is a dimensionless parameter $\lambda=\kappa\Lambda+1$ quantifying the effective cosmological constant at low curvature scales. Here we will assume $\Lambda=0$ since its contribution can be neglected in the early universe.}}
\begin{equation}
\mathcal{S}=\frac{1}{\kappa}\int d^4x\left(\sqrt{\left|g_{\mu\nu}+\kappa R_{(\mu\nu)}(\Gamma)\right|}-\sqrt{-g}\right)+\mathcal{S}_{\textrm{m}}\,,
\end{equation}
where $\kappa$ is the Born-Infeld coupling constant (we have assumed $8\pi G=c=1$). Note that $\kappa$ is dimensional and it can be either positive or negative. The EiBI theory is formulated with the Palatini variational principle in which the metric and the affine connection are treated independently at the Lagrangian level \cite{Banados:2010ix}. The Ricci tensor is constructed from the affine connection and only the symmetric part of $R_{\mu\nu}$ is considered. Due to the Born-Infeld structure (the square root) inherent in the theory, the big bang singularity can be avoided for a radiation dominated universe. Therefore, the description of the early universe is significantly different from that in GR. We expect that this would give rise to different physics of PBH formation. 

Assuming that the universe is dominated by a perfect fluid with energy density $\rho$ and a constant equation of state $w$, the modified Friedmann equation describing a closed universe can be written as
\begin{equation}
6\kappa H^2=X(w,\bar\rho)-\frac{6\kappa}{a^2}Y(w,\bar\rho)\,,\label{Friedeibi}
\end{equation}
where $H$ is the Hubble rate and 
\begin{align}
X(w,\bar\rho)&=\frac{1+2\sqrt{\frac{\left(1-w\bar\rho\right)^3}{1+\bar\rho}}-3\left(\frac{1-w\bar\rho}{1+\bar\rho}\right)}{\left[1+\frac{3}{4}\frac{\bar\rho(1+w)\left(2w\bar\rho+w-1\right)}{\left(1-w\bar\rho\right)\left(1+\bar\rho\right)}\right]^2}\,,\nonumber\\
Y(w,\bar\rho)&=\frac{\left(\frac{1-w\bar\rho}{1+\bar\rho}\right)}{\left[1+\frac{3}{4}\frac{\bar\rho(1+w)\left(2w\bar\rho+w-1\right)}{\left(1-w\bar\rho\right)\left(1+\bar\rho\right)}\right]^2}\,.\label{XYdef}
\end{align}
Here $\bar\rho\equiv\kappa\rho$ stands for the dimensionless energy density. Note that $\bar\rho$ can also be negative because $\kappa$ can be negative as well. When the energy density is small, i.e., $|\bar\rho|\ll1$, the Friedmann equation in GR is recovered:
\begin{equation}
H^2\approx\frac{\rho}{3}-\frac{1}{a^2}\qquad\textrm{when }|\bar\rho|\ll1\,.
\end{equation}
According to Eq.~\eqref{Friedeibi}, it can be shown that the evolution of the early universe in the EiBI gravity is quite different from that in GR. More explicitly, when the energy density is large, i.e., $|\bar\rho|=\mathcal{O}(1)$, the Born-Infeld corrections become significant. One of the important features in the EiBI gravity is its capability of removing the big bang singularity in a purely classical level, without any violation of energy conditions.  

In the picture of PBH formation, some mechanisms, such as inflation, are needed to generate initial quantum fluctuations seeding the density perturbations in the universe. The possibility of regarding the EiBI theory as an alternative to inflation has been discussed in Ref.~\cite{Avelino:2012ue}, in which the authors showed that the fundamental problems in standard cosmology, such as the flatness problem and the horizon problem may be solved in the EiBI gravity. Additionally, the quadratic inflationary paradigm in the EiBI gravity coupled with a scalar field (inflaton) has been proposed in \cite{Cho:2013pea} and studied in Refs.~\cite{Cho:2014ija,Cho:2014jta,Cho:2014xaa,Cho:2015yza,Cho:2015yua}. This inflationary model allows not only the generation of primordial quantum fluctuations we need for PBH formation, but also allows a well-defined pre-inflationary stage without the need of quantum gravity. Other interesting models  based on the Born-Infeld gravity which allow viable inflationary scenarios can be found in Refs.~\cite{Chen:2015eha,Jimenez:2015jqa}.

\subsection{PBH formation in EiBI gravity: Simple estimation}\label{subsec.1}
In order to estimate the threshold of density contrast for PBH formation, we assume that in the overdense region, the overdensity is uniform and is given by $\bar\rho(1+\delta)$, such that $\bar\rho$ stands for the background energy density. The modified Friedmann equation in the overdense region can be written as
\begin{equation}
6\kappa H^2=X^\delta-\frac{6\kappa}{a^2}Y^\delta\,,\label{expression1111}
\end{equation}
where we have defined $X^\delta\equiv X(w,\bar\rho(1+\delta))$ and $Y^\delta\equiv Y(w,\bar\rho(1+\delta))$ for the sake of abbreviation. 

Considering the time of horizon reentry and using the uniform Hubble slicing in Eq.~\eqref{expression1111}, one can obtain
\begin{equation}
\frac{Y^\delta_{\textrm{hc}}}{a^2_{\textrm{hc}}H^2_{\textrm{hc}}}=\frac{X^\delta_{\textrm{hc}}}{X_{\textrm{hc}}}-1\,.\label{interminhc}
\end{equation}
Combining it with Eq.~\eqref{reentryxa}, one can rewrite Eq.~\eqref{interminhc} as follows
\begin{equation}
\frac{1}{Y^\delta_{\textrm{hc}}}\left(\frac{X^\delta_{\textrm{hc}}}{X_{\textrm{hc}}}-1\right)=\sin^2\chi_a\,.
\label{2.10sin}
\end{equation}

As mentioned in \cite{Carr:1975qj}, the estimation of the threshold $\delta_{\textrm{th}}$ for PBH formation is based on the requirement that the proper size of the overdense region at the time of maximum expansion is larger than the Jeans length $R_\textrm{J}$. In Ref.~\cite{Carr:1975qj}, the Jeans length was approximated as $R_\textrm{J}\approx c_s\sqrt{3/\rho_\textrm{m}}\approx\sqrt{3w/\rho_\textrm{m}}$. However, in the EiBI gravity, the Jeans length has to be modified accordingly, as compared with that in GR. Let us, for the time being, denote the Jeans length in the EiBI gravity as $R_\textrm{J}^{\textrm{EiBI}}$. Following the estimation made in Ref.~\cite{Carr:1975qj} but adopting it in the EiBI gravity, the formation of PBHs would happen when
\begin{equation}
\left(R_\textrm{J}^{\textrm{EiBI}}\right)^2<a_\textrm{m}^2\sin^2\chi_a=\frac{a_\textrm{m}^2}{Y^\delta_{\textrm{hc}}}\left(\frac{X^\delta_{\textrm{hc}}}{X_{\textrm{hc}}}-1\right)\,,\label{jeansesti1}
\end{equation}
where Eq.~\eqref{2.10sin} has been used in the last equality. Since $a_\textrm{m}^2X_\textrm{m}=6\kappa Y_\textrm{m}$ at the maximum expansion time, the inequality \eqref{jeansesti1} can be further written as
\begin{equation}
\left(R_\textrm{J}^{\textrm{EiBI}}\right)^2<\frac{6\kappa Y_{\textrm{m}}}{Y^\delta_{\textrm{hc}}X_\textrm{m}}\left(\frac{X^\delta_{\textrm{hc}}}{X_{\textrm{hc}}}-1\right)\,.\label{jeansesti2}
\end{equation}

In the literature, the Jeans instability \cite{DeMartino:2017gye} and the corresponding Jeans length in the EiBI gravity has been investigated in the weak field limit, in which the modified Poisson equation can be written as \cite{Banados:2010ix}
\begin{equation}
\nabla^2\Phi=\frac{\rho}{2}+\frac{\kappa}{4}\nabla^2\rho\,,\label{modifeidPoisson}
\end{equation}
where $\Phi$ is the gravitational potential. Using Eq.~\eqref{modifeidPoisson}, the Jeans length of the EiBI gravity (in the weak field limit) can be obtained \cite{DeMartino:2017gye}:
\begin{equation}
R_\textrm{J}^{\textrm{EiBI}}=R_\textrm{J}\sqrt{1+\frac{\kappa}{2R_\textrm{J}^2}}\,,\label{Jeanseibi}
\end{equation}
where $R_\textrm{J}$ is its GR counterpart.

To proceed with the estimation, we will focus on the PBHs forming in the weak field regime, where $|\bar\rho|\ll1$, and expand the equation in terms of $\bar\rho$ up to linear order{\footnote{Note that Eq.~\eqref{jeansesti2} is an exact expression, without any approximation on $\bar\rho$ and $\delta$. The most challenging obstacle to proceed with the estimation is to derive the Jeans length of the EiBI gravity valid for arbitrary values of $\bar\rho$. This would be beyond the scope of this paper. However, we can still estimate the threshold in the weak field regime by using the Jeans length given by \eqref{Jeanseibi}. For another estimate, which is valid also in the strong field regime, see Sec.~\ref{subsec.2}.}. In a radiation dominated universe where $w=1/3$, we obtain
\begin{equation}
\delta^{\textrm{UH}}_{\textrm{hc}}>\delta_\textrm{th}\approx\frac{1}{3}\left(1+\frac{\bar\rho_\textrm{m}}{2}\right)\,,\label{threseibi13}
\end{equation}
where $\delta^{\textrm{UH}}_{\textrm{hc}}$ represents the density contrast at the horizon reentry in the uniform Hubble slicing. It can be seen that the threshold deviates from its GR counterpart $\delta_\textrm{th}=w=1/3$. In fact, for those PBHs which were generated not so early in the cosmic history such that the series expansion of $\bar\rho$ up to linear order is valid, the threshold would be enhanced (suppressed) if $\kappa$ is positive (negative), as compared with that in GR.} Furthermore, one can also see that the threshold depends on the background energy density at the time of PBH formation. This is due to the presence of the additional dimensional coupling constant $\kappa$ in the theory. 

Apparently, the estimation of $\delta_{\textrm{th}}$ here highly depends on the approximation of the Jeans length in Eq.~\eqref{jeansesti2}. In fact, as has been pointed out in Ref.~\cite{Harada:2013epa}, the choice of the Jeans length here could lead to some ambiguities about a numerical factor of order one. In GR, this is one of the reasons why the estimation of the threshold $\delta_\textrm{th}=w$ is far from consistent with that obtained from numerical simulations. Therefore, the threshold obtained in Eq.~\eqref{threseibi13} in the EiBI gravity is not expected to be accurate either. What we can really conclude in this stage is that the estimated threshold for PBH formation depends on the description of early universe, as well as on the underlying gravitational theories. Specifically, the threshold would depend on the energy scale at the time of PBH formation. This is expected not only in the EiBI theory, but also in gravitational theories with dimensional parameters corresponding to mass scales other than the Planck mass. In addition, we have to emphasized that the threshold obtained in \eqref{threseibi13} is valid only for PBH formation happening in the regime where the series expansion of $\bar\rho$ is valid up to linear order. For PBHs which were generated much earlier such that the Born-Infeld correction is significant ($|\bar\rho|=\mathcal{O}(1)$), this result is not valid anymore and a more careful estimate is needed. In the next subsection, we will estimate the threshold $\delta_\textrm{th}$ by using another method proposed in Ref.~\cite{Harada:2013epa}. In GR, this new analytic method has been shown to be more consistent qualitatively with the results obtained from numerical simulations. We will show that in this refined approach, the threshold indeed not only differs from its GR counterpart, but also changes with respect to the energy scale at the time of PBH formation. Furthermore, this method is not restricted within the weak field regime, as opposed to the result given in Eq.~\eqref{threseibi13}. In spite of this, this new method shares a qualitatively similar property with the result in \eqref{threseibi13} in the sense that the threshold is indeed enhanced (suppressed) when $\kappa$ is positive (negative).

\subsection{PBH formation in EiBI gravity: Refined estimation}\label{subsec.2}
As has been mentioned above, the estimation of the threshold $\delta_\textrm{th}$ for PBH formation carried out in \cite{Carr:1975qj}, and that in the previous subsection, relies on the choices of the Jeans length and on several additional assumptions. In Ref.~\cite{Harada:2013epa}, a more refined estimation of the threshold was proposed based on the comparison between the sound crossing distance and the proper size of the overdense region at the maximum expansion time. It was shown that for a spherically symmetric overdense region with uniform oversensity, the threshold is estimated as $\delta_{\textrm{th}}=\sin^2\left[\pi\sqrt{w}/(1+3w)\right]$ and it fits well qualitatively with that obtained from numerical simulations. 

In this subsection, we will apply this refined approach to estimate the threshold $\delta_\textrm{th}$ in the EiBI gravity. It should be noticed that the early universe in the EiBI gravity evolves very differently as compared with the standard big bang scenario in the sense that the big bang singularity can be replaced either with a primordial bouncing phase or a loitering phase, depending on the sign of the Born-Infeld coupling constant $\kappa$. When calculating the sound crossing distance, one has to integrate the infinitesimal distance from the beginning of the universe to the maximum expansion time. Therefore, the sound crossing distance could be significantly different in different cosmological scenarios, hence leading to an estimated threshold $\delta_\textrm{th}$ completely distinct from its GR counterpart. 

Similar to what we have done in the previous subsection, we first assume that the overdense region has uniform overdensity and can be described by a part of the closed Friedmann universe \eqref{closeFLRW}. The modified Friedmann equation in the EiBI gravity is given by Eq.~\eqref{Friedeibi}. In a radiation dominated universe where $w=c_s^2=1/3$, the comoving sound crossing distance is given by
\begin{align}
\int_0^{\chi_s}d\chi=\chi_s&=\sqrt{w}\int_{t_\textrm{i}}^{t_\textrm{m}}\frac{dt}{a}\nonumber\\
&=\frac{\sqrt{w}}{3}\int_{\bar\rho_\textrm{m}}^{\bar\rho_\textrm{i}}\frac{d\bar\rho}{\bar\rho(1+w)\sqrt{\left(\frac{\bar\rho_\textrm{m}}{\bar\rho}\right)^{\frac{2}{3(1+w)}}\frac{Y_\textrm{m}X}{X_\textrm{m}}-Y}}\,,\label{integrationchi}
\end{align}
where $X=X(w,\bar\rho)$ and $Y=Y(w,\bar\rho)$ are defined by Eq.~\eqref{XYdef}. The subscript ``i" represents the quantities at the time when the sound wave starts propagating. 

If the universe starts from a big bang singularity, the initial stage corresponds to $t_\textrm{i}=0$ and $\rho_\textrm{i}=\infty$. However, in the EiBI gravity, the energy density is bounded from above and the big bang singularity is replaced with either a bouncing scenario when $\kappa<0$ or a loitering stage when $\kappa>0$. For the bouncing scenario, the bounce happens when $\bar\rho=-1$ and it connects the contracting phase and the expanding phase of the universe. For the loitering scenario, the universe starts with a minimum size at the infinite past and the energy density is bounded by a maximum value $\bar\rho=1/w$. The integral in Eq.~\eqref{integrationchi} does not allow an analytic expression. Therefore, we will assume $w=1/3$ and obtain $\chi_s$ numerically. For the bouncing scenario, we will assume that the sound wave starts propagating at the bounce where $\bar\rho_\textrm{i}=-1$. As for the loitering scenario, on the other hand, we will assume that the sound wave starts propagating at a finite cosmic time in the past. In fact, the integration in Eq.~\eqref{integrationchi} does not converge if we assume $\bar\rho_\textrm{i}=1/w=3$. This means that the sound wave starts propagating in the infinite past and the wave can cross through an infinitely long distance. This does not make physical sense for a standard mechanism of PBH formation. Therefore, we will assume $\bar\rho_\textrm{i}=3-\Delta$, where $\Delta\gtrsim0$, when numerically calculating $\chi_s$.

After obtaining the sound crossing distance $\chi_s$, the threshold $\delta_\textrm{th}$ for PBH formation can be derived by requiring that the proper size of the overdense region at the time of maximum expansion is larger than the sound crossing distance, i.e., $\chi_a>\chi_s$. Using Eq.~\eqref{2.10sin}, we obtain
\begin{equation}
\frac{1}{Y^\delta_{\textrm{hc}}}\left(\frac{X^\delta_{\textrm{hc}}}{X_{\textrm{hc}}}-1\right)>\sin^2\chi_s\,.\label{refineresult}
\end{equation}
The scale factor $a$ and the energy density $\bar\rho\propto a^{-3(1+w)}$ appearing on the left-hand side of Eq.~\eqref{refineresult} are expressed at the time of horizon reentry. In order to see what these quantities would be at the maximum expansion time, we shall obtain the ratio of the energy density between the horizon reentry and the maximum expansion time, i.e., $\bar\rho_{\textrm{hc}}/\bar\rho_{\textrm{m}}$, for a given value of $\chi_a$. Therefore, we focus on the evolution of the energy density inside the overdense region and find that
\begin{align}
\sin^2\chi_a=\frac{1}{a_{\textrm{hc}}^2H_{\textrm{hc}}^2}&=\frac{6\kappa/a_{\textrm{m}}^2}{\left(\frac{a_{\textrm{hc}}}{a_{\textrm{m}}}\right)^2\left(X_{\textrm{hc}}-\frac{6\kappa}{a_{\textrm{hc}}^2}Y_{\textrm{hc}}\right)}\nonumber\\
&=\frac{X_\textrm{m}/Y_\textrm{m}}{\left(\frac{a_{\textrm{hc}}}{a_\textrm{m}}\right)^2X_\textrm{hc}-Y_\textrm{hc}\left(\frac{X_\textrm{m}}{Y_\textrm{m}}\right)}\,,\label{relation}
\end{align} 
where in the last equality, we have used the equation $X_{\textrm{m}}/Y_{\textrm{m}}=6\kappa/a_{\textrm{m}}^2$, which is obtained by the fact that $H_{\textrm{m}}=0$ at the maximum expansion time. In the calculation of the threshold for a given $\bar\rho_\textrm{m}$, one first derives $\chi_s$ by integrating \eqref{integrationchi}. Then, one inserts the value of $\chi_s$ into $\chi_a$, and calculates the ratio $\bar\rho_\textrm{hc}/\bar\rho_\textrm{m}$ by solving Eq.~\eqref{relation}. Finally, after replacing the $a_{\textrm{hc}}$ and $\bar\rho_{\textrm{hc}}$ on the left-hand side of the Eq.~\eqref{refineresult} in terms of $\bar\rho_{\textrm{m}}$, the threshold $\delta_{\textrm{th}}$ can be obtained numerically by solving Eq.~\eqref{refineresult}.
\begin{figure}
\includegraphics[scale=0.55]{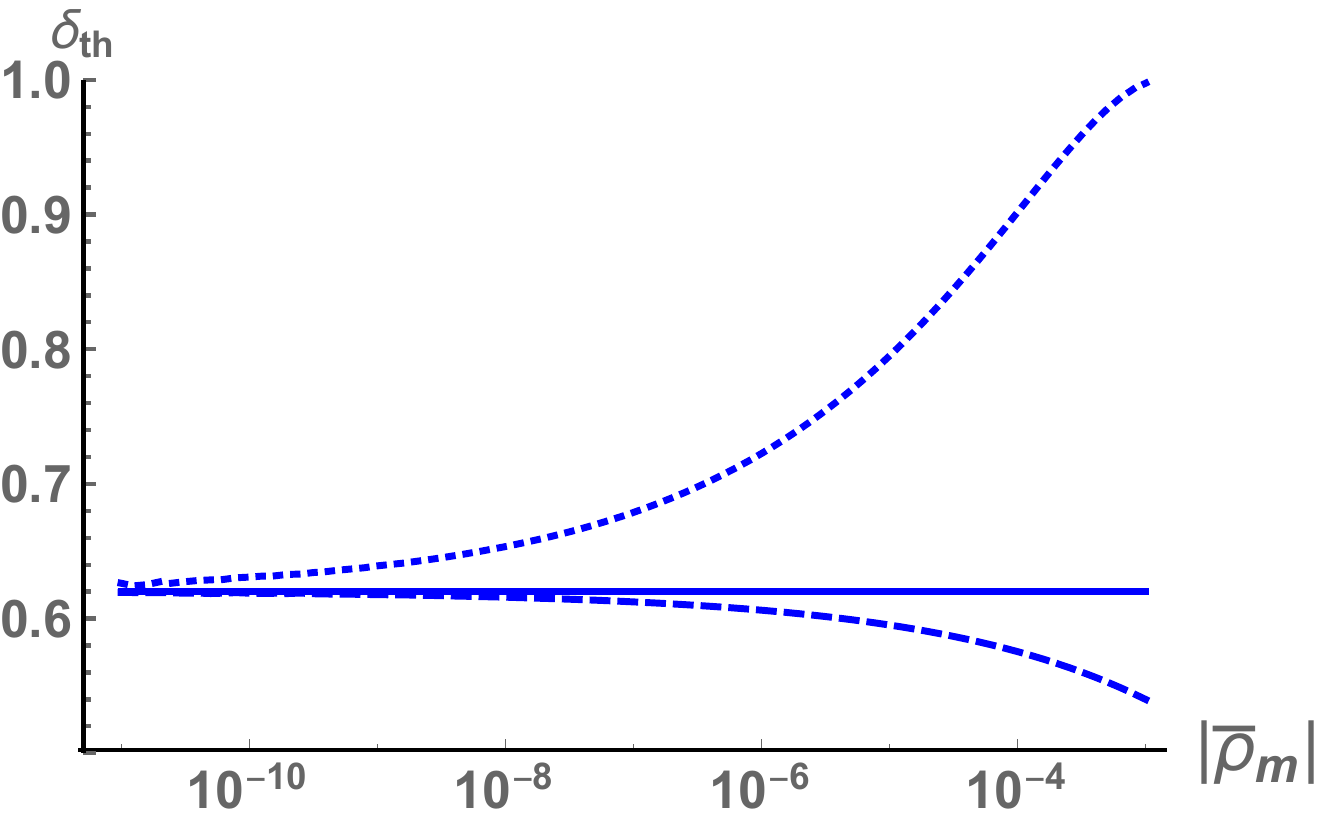}
\caption{\label{refinedres}The threshold $\delta_\textrm{th}$ of the density contrast for PBH formation estimated by using the refined approach \cite{Harada:2013epa}. The dashed (dotted) curve corresponds to a negative (positive) $\kappa$. We have assumed $\Delta=10^{-6}$ in the case of positive $\kappa$. The horizontal solid line represents the threshold value $\sin^2\left(\pi/2\sqrt{3}\right)=0.62$ obtained in GR. It can be seen that in the EiBI gravity, the threshold estimated based on this approach depends on the energy scale at the time of PBH formation.}
\end{figure}

In FIG.~\ref{refinedres}, we show the threshold $\delta_\textrm{th}$ of the density contrast for PBH formation estimated in the EiBI gravity. The dashed (dotted) curve corresponds to a negative (positive) Born-Infeld coupling constant $\kappa$. For the case of positive $\kappa$, we have assumed $\Delta=10^{-6}$. It can be shown that if the value of $\Delta$ is smaller, the sound wave starts propagating earlier and the threshold for PBH formation would be larger. In addition, the threshold would be enhanced (suppressed) when the Born-Infeld coupling constant $\kappa$ is positive (negative). This is qualitatively consistent with the result of the previous analytic estimation \eqref{threseibi13}.

Although the two approaches of threshold estimation in Sec.~\ref{subsec.1} and Sec.~\ref{subsec.2} give rise to different quantitative results, they share similar qualitative properties. For example, it can be seen that for both approaches, the threshold estimated in the EiBI gravity not only depends on the sign of the Born-Infeld coupling constant $\kappa$, it also depends on the energy scale $\bar\rho_\textrm{m}$ at the time of PBH formation. If the density contrast starts to collapse quite late such that the background energy density is low enough at the onset of PBH formation ($|\bar\rho_\textrm{m}|\ll1$), the threshold value in GR can be recovered. On the other hand, if the density contrast starts collapsing in an early time in which the deviation between GR and EiBI gravity is significant, the threshold could be significantly different from its GR counterpart. We would like to emphasize that in the second approach of estimation, we do not assume $|\bar\rho|\ll1$, as opposed to the first approach. In fact, the threshold estimated in the EiBI gravity has already differed from that in GR significantly when $|\bar\rho|=\mathcal{O}(10^{-4})$. This can be easily seen from FIG.~\ref{refinedres}.

\section{PBH mass function and abundance}\label{sec.mfa}
In the theory of PBH formation, the value of the threshold $\delta_\textrm{th}$ plays a very important role in determining the mass function as well as the abundance of PBHs. One of the important physical quantities is the fraction $\beta(M)$ of the universe, which forms PBHs at a certain time, with their initial mass larger than a chosen $M$. In the literature, the estimation of this fraction is commonly carried our by applying the Press-Schecher formalism \cite{Press:1973iz}. In this formalism, the fraction $\beta(M)$ can be written as
\begin{equation}
\beta(M)=\int^\infty_{\delta_{\textrm{c}}(M)}P(\delta(M))d\delta(M)\,,
\end{equation}  
where $\delta(M)$ is the smoothed density contrast defined by the convolution between the density contrast and an appropriate window function. Usually, these quantities are defined on a comoving slicing. The lower bound of the integral $\delta_{\textrm{c}}(M)$ corresponds to the formation threshold of the density contrast on the same slicing. Generically, the threshold $\delta_{\textrm{c}}(M)$ can depend on the mass scale under consideration. Furthermore, $P(\delta(M))$ is the probability density distribution of the density contrast.   

Assuming the probability density distribution to be gaussian, the fraction $\beta(M)$ can be written as
\begin{align}
\beta(M)&=\int^\infty_{\delta_{\textrm{c}}(M)}\frac{2}{\sqrt{2\pi\sigma^2(M)}}\,\textrm{exp}\left(-\frac{\delta^2(M)}{2\sigma^2(M)}\right)d\delta(M)\nonumber\\
&\approx\sqrt{\frac{2\sigma^2(M)}{\pi\delta_\textrm{c}^2(M)}}\,\textrm{exp}\left(-\frac{\delta_\textrm{c}^2(M)}{2\sigma^2(M)}\right)\,,\label{betam}
\end{align} 
where $\sigma(M)$ is the variance of $\delta(M)$. In the second line of Eq.~\eqref{betam}, we have assumed $\delta_{\textrm{c}}(M)\gg\sigma(M)$ in the case of interest. In principle, the estimation would give different results if we consider different primordial curvature power spectra. Note that the primordial curvature power spectra depend on the variance. The most important point at this stage is that the fraction $\beta(M)$ is highly sensitive to the change of $\delta_{\textrm{c}}$, because of the exponential dependence, i.e., the $\textrm{exp}\left(-\delta_\textrm{c}^2/2\sigma^2\right)$ factor (note that we have $\sigma\ll1$). Even a small amount of changes of the formation threshold $\delta_{\textrm{c}}$ could largely alter the estimation of the fraction $\beta(M)$. This can be seen, for example, in Figure 4 of Ref.~\cite{Young:2014ana}. Therefore, the possibility that the threshold value depends on the underlying gravitational theory should not be overlooked. Indeed, according to FIG.~\ref{refinedres}, the threshold of PBH formation can deviate from its GR counterpart up to $30\%$, even when $|\bar\rho|$ is only $10^{-4}$. This deviation, when propagating to the deviation of $\beta(M)$, would be largely amplified.

In fact, the estimated fraction $\beta(M)$ would also depend on which formalism we are choosing to estimate $\beta(M)$. If one considers the peaks theory \cite{Bardeen:1985tr}, rather than the Press-Schecher formalism, to do the estimation, the result would be different, although it has been shown that the result is less sensitive to the use of different formalisms, compared with the change of the threshold value \cite{Green:2004wb,Young:2014ana}. It should be mentioned that the peak theory has recently been accepted as a more accurate approach compared with the Press-Schecher formalism \cite{Germani:2018jgr}, partially because of the inaccuracy of the gaussian assumption made in the latter \cite{Yoo:2018kvb}. However, the conclusion that the abundance highly depends on the threshold value is true for both formalisms. Also, it should be emphasized that the peaks theory and the Press-Schecher formalism have both assumed that PBHs with different masses are formed at the same time. However, in principle PBHs with different masses correspond to different moments of horizon reentry, hence different formation time. In modified theories of gravity, as we have shown, the threshold $\delta_\textrm{c}$ would even depend on the background energy scale at the time of PBH formation (even considering the top-hat profile). In order to have a more accurate estimation, therefore, some revised formalisms should be taken into account. See Ref.~\cite{Suyama:2019npc} for a recent related development.

\section{Conclusions}\label{sec.conclu}
In this paper, we show explicitly that the threshold of density contrast for PBH formation depends on the gravitational theories under consideration. The estimated value of the threshold differs from its GR counterpart when considering gravitational theories which modify the early time description of the universe. The threshold estimations can be performed analytically by assuming that the overdensity is homogeneous in the overdense regions eventually collapsing to form PBHs. In GR, given a constant equation of state $w$ in the background universe, the analytic estimations give a constant threshold depending only on $w$. However, in modified theories of gravity with additional characteristic mass scales, the threshold would not only deviate from its GR counterpart, but also depend on the energy scale at the time of PBH formation. In the EiBI gravity, we indeed find that the threshold depends on the energy scale at the time of PBH formation. Also, the threshold would be enhanced (suppressed) when the Born-Infeld coupling constant $\kappa$ is positive (negative). Although we have only considered the EiBI gravity in this paper, we expect that similar conclusions that the threshold depends on the energy scale at the PBH formation time can be drawn in other gravitational theories with additional mass scales.

Since the threshold of density contrast for PBH formation plays a very crucial role in determining the mass function and abundance of PBHs, we can conclude that the changes of gravitational theories and inflationary models would affect the PBH mass function not just through the primordial power spectrum, they may modify the threshold value and largely change the predictions of PBH mass function. It can be seen that the prediction of PBH mass function depends on several physical mechanisms, which implies that there would be huge degeneracy in the parameter space. However, if there are other observational constraints which are able to break this degeneracy, it is expected that PBH physics could be another powerful tool to test gravitational theories in the future. For example, the Born-Infeld coupling constant $\kappa$ in the EiBI gravity is able to be constrained from solar system \cite{Casanellas:2011kf}, cosmology \cite{Avelino:2012ge}, nuclear physics \cite{Avelino:2012qe,Avelino:2019esh}, and recently from the speed of gravitational waves \cite{Jana:2017ost}. This is similar to the ability of PBH physics in probing the dynamics of our early universe.

It should be emphasized that in this paper we only consider the analytic estimations of the threshold by assuming uniform overdensity, i.e., the top-hat profile. It is in fact well-known that the threshold value would also depend on the density profile under consideration \cite{Nakama:2013ica,Germani:2018jgr,Musco:2018rwt,Young:2019osy,Escriva:2019phb,Germani:2019zez}. In these scenarios, analytic estimation is not feasible anymore and one shall resort to numerical simulations. It would be interesting to implement a more rigorous threshold estimation in the framework of modified theories of gravity considering general overdensity profiles. We leave these issues for our future works.

\section*{Acknowledgement}

The author would like to thank the anonymous referee for fruitful and important comments about this paper. CYC is supported by Ministry of Science and Technology (MOST), Taiwan, through No. 107-2119-M-002-005 and No. 108-2811-M-002-682. He is also supported by Leung Center for Cosmology and Particle Astrophysics (LeCosPA) of National Taiwan University, Taiwan National Center for Theoretical Sciences (NCTS), and Institute of Physics of Academia Sinica.


\begin{thebibliography}{99}

\bibitem{Hawking:1971ei} 
  S.~Hawking,
  Mon.\ Not.\ Roy.\ Astron.\ Soc.\  {\bf 152}, 75 (1971).

\bibitem{Polnarev:1986bi} 
  A.~G.~Polnarev and M.~Y.~Khlopov,
  Sov.\ Phys.\ Usp.\  {\bf 28}, 213 (1985)
  [Usp.\ Fiz.\ Nauk {\bf 145}, 369 (1985)].

\bibitem{Carr:2005zd} 
  B.~J.~Carr,
  astro-ph/0511743.


\bibitem{Khlopov:1980mg} 
  M.~Y.~Khlopov and A.~G.~Polnarev,
  Phys.\ Lett.\  {\bf 97B}, 383 (1980).

\bibitem{Kalashnikov:1983qv} 
  O.~K.~Kalashnikov and M.~Y.~Khlopov,
  Phys.\ Lett.\  {\bf 127B}, 407 (1983).

\bibitem{Belotsky:2014kca} 
  K.~M.~Belotsky {\it et al.},
  Mod.\ Phys.\ Lett.\ A {\bf 29}, no. 37, 1440005 (2014).

\bibitem{Carr:2016drx} 
  B.~Carr, F.~Kuhnel and M.~Sandstad,
  Phys.\ Rev.\ D {\bf 94}, no. 8, 083504 (2016).

\bibitem{Inomata:2017okj} 
  K.~Inomata, M.~Kawasaki, K.~Mukaida, Y.~Tada and T.~T.~Yanagida,
  Phys.\ Rev.\ D {\bf 96}, no. 4, 043504 (2017).

\bibitem{Ballesteros:2017fsr} 
  G.~Ballesteros and M.~Taoso,
  Phys.\ Rev.\ D {\bf 97}, no. 2, 023501 (2018).

\bibitem{Belotsky:2018wph} 
  K.~M.~Belotsky {\it et al.},
  Eur.\ Phys.\ J.\ C {\bf 79}, no. 3, 246 (2019).

\bibitem{Sasaki:2018dmp} 
  M.~Sasaki, T.~Suyama, T.~Tanaka and S.~Yokoyama,
  Class.\ Quant.\ Grav.\  {\bf 35}, no. 6, 063001 (2018).



\bibitem{Konoplich:1999qq} 
  R.~V.~Konoplich, S.~G.~Rubin, A.~S.~Sakharov and M.~Y.~Khlopov,
  Phys.\ Atom.\ Nucl.\  {\bf 62}, 1593 (1999)
  [Yad.\ Fiz.\  {\bf 62}, 1705 (1999)].

\bibitem{Rubin:2000dq} 
  S.~G.~Rubin, M.~Y.~Khlopov and A.~S.~Sakharov,
  Grav.\ Cosmol.\  {\bf 6}, 51 (2000).

\bibitem{Khlopov:2000js} 
  M.~Y.~Khlopov, R.~V.~Konoplich, S.~G.~Rubin and A.~S.~Sakharov,
  Grav.\ Cosmol.\  {\bf 6}, 153 (2000).


\bibitem{Rubin:2001yw} 
  S.~G.~Rubin, A.~S.~Sakharov and M.~Y.~Khlopov,
  J.\ Exp.\ Theor.\ Phys.\  {\bf 91}, 921 (2001)
  [J.\ Exp.\ Theor.\ Phys.\  {\bf 92}, 921 (2001)].

\bibitem{Khlopov:2004sc} 
  M.~Y.~Khlopov, S.~G.~Rubin and A.~S.~Sakharov,
  Astropart.\ Phys.\  {\bf 23}, 265 (2005).

\bibitem{Khlopov:2008qy} 
  M.~Y.~Khlopov,
  Res.\ Astron.\ Astrophys.\  {\bf 10}, 495 (2010).

\bibitem{Carr:2009jm}
B.~Carr, K.~Kohri, Y.~Sendouda and J.~Yokoyama,
Phys. Rev. D \textbf{81}, 104019 (2010).

\bibitem{Niemeyer:1999ak} 
  J.~C.~Niemeyer and K.~Jedamzik,
  Phys.\ Rev.\ D {\bf 59}, 124013 (1999).

\bibitem{Shibata:1999zs} 
  M.~Shibata and M.~Sasaki,
  Phys.\ Rev.\ D {\bf 60}, 084002 (1999).

\bibitem{Musco:2004ak} 
  I.~Musco, J.~C.~Miller and L.~Rezzolla,
  Class.\ Quant.\ Grav.\  {\bf 22}, 1405 (2005).

\bibitem{Polnarev:2006aa} 
  A.~G.~Polnarev and I.~Musco,
  Class.\ Quant.\ Grav.\  {\bf 24}, 1405 (2007).

\bibitem{Musco:2008hv} 
  I.~Musco, J.~C.~Miller and A.~G.~Polnarev,
  Class.\ Quant.\ Grav.\  {\bf 26}, 235001 (2009).

\bibitem{Musco:2012au} 
  I.~Musco and J.~C.~Miller,
  Class.\ Quant.\ Grav.\  {\bf 30}, 145009 (2013).

\bibitem{Young:2019yug} 
  S.~Young, I.~Musco and C.~T.~Byrnes,
  JCAP {\bf 1911}, no. 11, 012 (2019).


\bibitem{Escriva:2019nsa}
A.~Escriv\`a,
Phys. Dark Univ. \textbf{27}, 100466 (2020).



\bibitem{Carr:1975qj} 
  B.~J.~Carr,
  Astrophys.\ J.\  {\bf 201}, 1 (1975).
  
\bibitem{Harada:2013epa} 
  T.~Harada, C.~M.~Yoo and K.~Kohri,
  Phys.\ Rev.\ D {\bf 88}, no. 8, 084051 (2013)
  Erratum: [Phys.\ Rev.\ D {\bf 89}, no. 2, 029903 (2014)].

\bibitem{Nakama:2013ica} 
  T.~Nakama, T.~Harada, A.~G.~Polnarev and J.~Yokoyama,
  JCAP {\bf 1401}, 037 (2014).

\bibitem{Germani:2018jgr} 
  C.~Germani and I.~Musco,
  Phys.\ Rev.\ Lett.\  {\bf 122}, no. 14, 141302 (2019).

\bibitem{Musco:2018rwt} 
  I.~Musco,
  Phys.\ Rev.\ D {\bf 100}, no. 12, 123524 (2019).


\bibitem{Young:2019osy}
S.~Young,
Int. J. Mod. Phys. D \textbf{29}, no.02, 2030002 (2019).



\bibitem{Escriva:2019phb}
A.~Escriv\`a, C.~Germani and R.~K.~Sheth,
Phys. Rev. D \textbf{101}, no.4, 044022 (2020).


\bibitem{Germani:2019zez}
C.~Germani and R.~K.~Sheth,
Phys. Rev. D \textbf{101}, no.6, 063520 (2020).

\bibitem{Harada:2017fjm} 
  T.~Harada, C.~M.~Yoo, K.~Kohri and K.~I.~Nakao,
  Phys.\ Rev.\ D {\bf 96}, no. 8, 083517 (2017)
  Erratum: [Phys.\ Rev.\ D {\bf 99}, no. 6, 069904 (2019)].

\bibitem{He:2019cdb} 
  M.~He and T.~Suyama,
  Phys.\ Rev.\ D {\bf 100}, no. 6, 063520 (2019).

\bibitem{Fernandez:2019kyb}
N.~Fernandez and S.~Profumo,
JCAP \textbf{08}, 022 (2019).

\bibitem{Khlopov:1985jw} 
  M.~Khlopov, B.~A.~Malomed and I.~B.~Zeldovich,
  Mon.\ Not.\ Roy.\ Astron.\ Soc.\  {\bf 215}, 575 (1985).

\bibitem{Hidalgo:2017dfp} 
  J.~C.~Hidalgo, J.~De Santiago, G.~German, N.~Barbosa-Cendejas and W.~Ruiz-Luna,
  Phys.\ Rev.\ D {\bf 96}, no. 6, 063504 (2017).

\bibitem{Capozziello:2011et} 
  S.~Capozziello and M.~De Laurentis,
  Phys.\ Rept.\  {\bf 509}, 167 (2011).

\bibitem{Nojiri:2017ncd} 
  S.~Nojiri, S.~D.~Odintsov and V.~K.~Oikonomou,
  Phys.\ Rept.\  {\bf 692}, 1 (2017).

\bibitem{Chen:2016kjx} 
  J.~W.~Chen, J.~Liu, H.~L.~Xu and Y.~F.~Cai,
  Phys.\ Lett.\ B {\bf 769}, 561 (2017).

\bibitem{Bhattacharya:2019bvk} 
  S.~Bhattacharya, S.~Mohanty and P.~Parashari,
  arXiv:1912.01653 [astro-ph.CO].

\bibitem{Dialektopoulos:2017pgo} 
  K.~F.~Dialektopoulos, A.~Nathanail and A.~G.~Tzikas,
  Phys.\ Rev.\ D {\bf 97}, no. 12, 124059 (2018).

\bibitem{Vallejo-Pena:2019hgv} 
  S.~A.~Vallejo-Pe\~na and A.~E.~Romano,
  arXiv:1911.03327 [gr-qc].

\bibitem{Banados:2010ix} 
  M.~Ba\~nados and P.~G.~Ferreira,
  Phys.\ Rev.\ Lett.\  {\bf 105}, 011101 (2010)
  Erratum: [Phys.\ Rev.\ Lett.\  {\bf 113}, no. 11, 119901(E) (2014)].
    
\bibitem{BeltranJimenez:2017doy} 
  J.~Beltr\'an Jim\'enez, L.~Heisenberg, G.~J.~Olmo and D.~Rubiera-Garcia,
  Phys.\ Rept.\  {\bf 727}, 1 (2018).
    
\bibitem{Scargill:2012kg}
  J.~H.~C.~Scargill, M.~Ba\~{n}ados and P.~G.~Ferreira,
  Phys.\ Rev.\ D {\bf 86} (2012) 103533.

\bibitem{Avelino:2012ue} 
  P.~P.~Avelino and R.~Z.~Ferreira,
  Phys.\ Rev.\ D {\bf 86}, 041501 (2012).

\bibitem{Cho:2013pea} 
  I.~Cho, H.~C.~Kim and T.~Moon,
  Phys.\ Rev.\ Lett.\  {\bf 111}, 071301 (2013).

\bibitem{Cho:2014ija} 
  I.~Cho and H.~C.~Kim,
  Phys.\ Rev.\ D {\bf 90}, no. 2, 024063 (2014).

\bibitem{Cho:2014jta} 
  I.~Cho and N.~K.~Singh,
  Eur.\ Phys.\ J.\ C {\bf 74}, no. 11, 3155 (2014).

\bibitem{Cho:2014xaa} 
  I.~Cho and N.~K.~Singh,
  Eur.\ Phys.\ J.\ C {\bf 75}, no. 6, 240 (2015).

\bibitem{Cho:2015yza} 
  I.~Cho and N.~K.~Singh,
  Phys.\ Rev.\ D {\bf 92}, no. 2, 024038 (2015).

\bibitem{Cho:2015yua} 
  I.~Cho and J.~O.~Gong,
  Phys.\ Rev.\ D {\bf 92}, no. 6, 064046 (2015).

\bibitem{Chen:2015eha} 
  C.~Y.~Chen, M.~Bouhmadi-L\'opez and P.~Chen,
  Eur.\ Phys.\ J.\ C {\bf 76}, 40 (2016).

\bibitem{Jimenez:2015jqa} 
  J.~Beltr\'an Jim\'enez, L.~Heisenberg, G.~J.~Olmo and C.~Ringeval,
  JCAP {\bf 1511}, 046 (2015).

\bibitem{DeMartino:2017gye}
I.~Martino and A.~Capolupo,
Eur. Phys. J. C \textbf{77}, no.10, 715 (2017).
 
 
\bibitem{Press:1973iz} 
  W.~H.~Press and P.~Schechter,
  Astrophys.\ J.\  {\bf 187}, 425 (1974).
  
  
\bibitem{Young:2014ana} 
  S.~Young, C.~T.~Byrnes and M.~Sasaki,
  JCAP {\bf 1407}, 045 (2014).


\bibitem{Bardeen:1985tr} 
  J.~M.~Bardeen, J.~R.~Bond, N.~Kaiser and A.~S.~Szalay,
  Astrophys.\ J.\  {\bf 304}, 15 (1986).
    
\bibitem{Green:2004wb} 
  A.~M.~Green, A.~R.~Liddle, K.~A.~Malik and M.~Sasaki,
  Phys.\ Rev.\ D {\bf 70}, 041502 (2004).
    
    
\bibitem{Yoo:2018kvb}
C.~M.~Yoo, T.~Harada, J.~Garriga and K.~Kohri,
PTEP \textbf{2018}, no.12, 123E01 (2018).
  
  
    
\bibitem{Suyama:2019npc}
T.~Suyama and S.~Yokoyama,
PTEP \textbf{2020}, no.2, 023E03 (2020).
    
    
    
\bibitem{Casanellas:2011kf} 
  J.~Casanellas, P.~Pani, I.~Lopes and V.~Cardoso,
  Astrophys.\ J.\  {\bf 745}, 15 (2012).
    
    
\bibitem{Avelino:2012ge} 
  P.~P.~Avelino,
  Phys.\ Rev.\ D {\bf 85}, 104053 (2012).
    
\bibitem{Avelino:2012qe} 
  P.~P.~Avelino,
  JCAP {\bf 1211}, 022 (2012).
    
\bibitem{Avelino:2019esh} 
  P.~P.~Avelino,
  Phys.\ Lett.\ B {\bf 795}, 627 (2019).
    
\bibitem{Jana:2017ost} 
  S.~Jana, G.~K.~Chakravarty and S.~Mohanty,
  Phys.\ Rev.\ D {\bf 97}, no. 8, 084011 (2018).
  
  
  
  
 

\end{thebibliography}
\end{document}